\documentstyle[12pt]{article}
\def\be{\begin{equation}}
\def\ee{\end{equation}}

\begin{document}
\baselineskip=18pt
\pagestyle{empty}
\begin{center}
\bigskip

\rightline{CWRU-P1-99}

\vskip 0.7in

{\bf \Large Life, The Universe, and Nothing:}
\vskip 0.2in

{\bf \Large Life and Death in an Ever-Expanding Universe}

\vskip 0.7in

Lawrence M. Krauss and Glenn D. Starkman 

\vskip 0.1in
{\it Departments of Physics and Astronomy\\
Case Western Reserve University \\
10900 Euclid Ave, Cleveland OH 44106-7079}

\end{center}

\vskip 0.8in

\begin{abstract} Current evidence suggests that
the cosmological constant is not zero, or that we live in an
open universe. We examine the implications for the future under
these assumptions, and find that they are striking. 
If the Universe is cosmological constant-dominated, 
our ability to probe the evolution of large scale structure 
will decrease with time ---
presently observable distant sources will disappear 
on a time-scale comparable to the period of stellar burning. 
Moreover, while the Universe might expand forever, 
the integrated conscious lifetime of any civilization will be finite,
although it can be astronomically long.   
We find that this latter result is far more general.  
In the absence of possible exotic and uncertain
strong gravitational effects, 
the total information recoverable by any civilization  over the 
entire history of our universe is finite,
and  assuming that consciousness has a physical computational basis,
life cannot be eternal.

\end{abstract}

\newpage
\pagestyle{plain}
\setcounter{page}{1}

Our universe could end in one of two ways.
Either the observed expansion could terminate 
and be followed by collapse and a Big Crunch 
or the expansion could continue forever.  The evidence
is overwhelmingly in favor of the latter possibility.
Indeed recent direct \cite{perl,kirsh} 
and indirect \cite{kraussturner,ostriker,krauss} 
measurements suggest that the expansion is { \it accelerating}, implying
that it is  driven by an energy density which at least mimics vacuum
energy, a so-called  cosmological constant.

As dramatic as this result may be for our understanding of fundamental
processes underlying the Big Bang, 
it has equally important consequences for the long-term 
quality of life of any conscious beings 
that may survive the more mundane challenges of daily existence. 
In an eternally-expanding universe life might,
at least in principle, endure forever \cite{dyson}. 
While global warming, nuclear war and asteroid impacts
may currently threaten human civilization, 
one may hope that humanity will overcome these threats, 
expand into the Universe,
and perhaps even encounter other intelligent life-forms. 
In any case, if intelligent life is ubiquitous in the Universe, 
it is reasonable to expect that no local threats can ever
wipe the slate entirely clean. 

But are there global constraints on the perdurability
or on the quality of conscious life in our Universe?  
These are the questions we examine here.   

We find that the future is particularly discouraging 
if we live in a cosmological-constant-dominated universe.
In this case, very soon, on a cosmic time-scale,
our ability to gather information on 
the large scale structure of the universe 
will begin to forever {\it decrease}.
The decreasing information base in the observable universe
is associated with a finite and decreasing supply of accessible 
energy.

Life's long term prospects 
are only slightly less dismal in any other cosmology, however.
We argue that the total energy that any civilization 
can ever recover and metabolize is finite, 
as is the recoverable information content,
independent of the geometry or expansion history of the universe.

Faced with this inevitable long term energy crisis,
life must eventually identify a strategy for reduced energy consumption  
or cease to exist. 
In a cosmological-constant dominated universe,
the de Sitter temperature fixes a minimum temperature below which
life cannot operate without energy-consuming refrigerators.
In any cosmology, the need to dissipate excess heat 
may fix a minimum temperature 
at which a biological system can operate continuously.

A minimum temperature in a  biological system 
of fixed information-theoretic complexity 
implies a minimum metabolic rate.
Faced with a minimum rate of energy consumption and a finite energy supply,
increasingly long hibernation seems the obvious alternative.
But, this requires perfectly reliable alarm clocks.
Statistically all alarm clocks eventually fail.
Furthermore, alarm clocks operating in thermal backgrounds
have minimum power consumption requirements.  
The options:  live for the moment in high-powered luxury, or
progressively reduce the information theoretic complexity of
life until it loses consciousness forever.  

The only remaining hope involves (almost) dissipationless computation.  
Under certain assumptions about the rate at which systems could
in principle dissipate the heat generated during such computation, 
it is possible to find a mathematical solution 
allowing an infinite  number of computations with finite energy.  
However, with a finite supply of information 
only a finite number of these computations are distinct.
Moreover, even if one accepts the reduction of consciousness to computation,
the generic features of physical consciousness  necessitate dissipation --  
namely observation, 
and, for a system of necessarily finite memory capacity,
 the erasure of inessential  memories.
We argue that these features imply that no finite system 
can perform an infinite number of computations with finite energy. 
Thus only a finite (if still huge) stream of 
consciousness is 
available to any civilization.

\section{Knowledge Decreases with Time}

George Orwell wrote, ``To see what is in front of one's nose 
requires a constant struggle."  
If the universe is dominated by a cosmological constant
this will become more true, with a vengeance, as time proceeds. 

The observable universe 
is remarkably homogeneous and isotropic on large scales. 
These properties enable us to parametrize the
evolution of  the universe's large scale geometry in terms
of one spatially homogeneous function of time, the scale factor $a(t)$.
The observed expansion of the universe can be understood as 
the increase in $a(t)$.
For objects comoving with this expansion,
$a(t)$ describes how the distance between them changes.
The evolution of the scale factor is given by the 
Einstein field equation appropriate for our very symmetric universe,
the Lemaitre-Friedmann-Robertson-Walker (LFRW) equation:
\be
\left({{\dot a}\over a}\right)^2 + {k\over a^2} = {8\pi G\rho\over 3} .
\label{FRWeqns}
\ee
Here $G$ is Newton's constant, $\rho$ is the energy density, 
and $k$ measures the curvature of space.
The expansion  history $a(t)$ depends strongly  
on:\hfil\break\noindent
(1) the  sign of $k$; \hfil\break\noindent
(2) the dependence of $\rho$ on $a$, 
in particular the $a$-dependence of 
the most slowly varying component of the density.  
For all known equations-of-state,
the time derivatives of $\rho$  and $a$ have the same sign.

If the universe becomes dominated  by a constant positive energy density 
$ \rho_\Lambda\equiv{\Lambda \over 8 \pi G}$,
then the evolution of the metric quickly approaches that
associated with a flat ($k=0$) Einstein-de Sitter universe, 
in which 
\be
a(t) = a(t_o)e^{\sqrt{\Lambda\over 3}(t-t_o)} .
\label{FRWeqns2}
\ee
$\Lambda$ is called the cosmological-constant, 
and $\rho_{\Lambda}$ may be interpreted as the intrinsic energy density 
associated with the vacuum.

From equation (\ref{FRWeqns2}), 
a point initially a distance $d$ away from an observer in such a universe
will be carried away by the cosmic expansion at a velocity
\be
{\dot d}  =\sqrt{{\Lambda \over 3}} \  d .
\ee
Equating this recession velocity to the speed of light $c$, 
one finds the physical distance to the so-called de Sitter horizon
as measured by a network of observers comoving with the expansion.  
This horizon, 
is a sphere enclosing a region, outside of which no new
information can reach the observer at the center, and across which
the outward de Sitter expansion carries material.  
Each observer has such a horizon sphere centered on them. 
Similarly, any signal we send out today will never reach
objects currently located distances further than the horizon
distance.  Moreover, this  distance may  be 
comparable to the current observable
region of the universe.
If we accept a cosmological constant of the magnitude
suggested by the current data, 
then $\rho_\Lambda \simeq 6 \times 10^{-30}$gm/cm$^3$ and 
the distance to the horizon is approximately 
$R_H\simeq1.7 \times 10^{26}{\rm m} \simeq 18$ billion light 
years.    

While the effects of the de Sitter horizon are not yet directly discernible, 
this result suggests that they will be seen on a time-scale 
comparable to the present age of the universe.  
As objects approach the horizon, 
the time (as measured by the clocks of the comoving observers) 
between the emission of light 
and its reception on Earth grows exponentially.
As the light travels from its source to the observer, 
its wavelength is stretched in proportion to the growth in $a(t)$.
Objects therefore appear exponentially redshifted as they 
approach the horizon.
Finally, their apparent brightness declines exponentially, 
so that the distance of the objects inferred by an
observer increases exponentially.
While it strictly takes an infinite amount of time
for the observer to completely
lose causal contact with these receding objects,
distant stars, galaxies, and all radiation backgrounds from the Big Bang 
will effectively  ``blink" out of existence in a finite time -- 
as their signals redshift,
the time scale for detecting these signals becomes
comparable to the age of the universe, as we describe below.

Eventually all objects not decoupled from the background 
expansion, 
{\it i.e.}~ those objects not bound to the local supercluster, 
will disappear in this fashion.  
The time-scale for this disappearance is surprisingly short.   
We can estimate it by taking a radius of 
$R_{SC}=10$ Megaparsecs (about $3\times10^7$ light years, 
$3\times10^{22}$m)
as the extent of the local supercluster of galaxies --
the largest observed structure of which we are a part. 
Objects further than this distance now will reach an
apparent distance $R_H$ in a time given by
\be 
{R_H \over R_{SC}} \simeq{1.7 \times 10^{26} m \over 3 \times 
10^{22} m} 
\simeq 5 \times 10^3 = \exp\left[\sqrt{\Lambda \over 3} t \right].
\ee
Thus, in roughly 150 billion years
light from all objects outside our local supercluster will
have redshifted by more than a factor of $5000$,
with each successive 150 billion years bringing an equal redshift factor.
In a little more than two trillion years, 
all extra-supercluster objects will have redshifted
by a factor of more than $10^{53}$.
Even for the highest energy gamma rays, 
a redshift of $10^{53}$ stretches 
their wavelength to greater than the physical diameter of the horizon.
(There is no contradiction here.  From the point of view of
a comoving observer, the horizon appears infinitely far away.  
Infinitely large redshift means that objects possessing such redshifts 
will have expanded infinitely far away by the time their light arrives
at the observer.)  The resolution time for such radiation will exceed
the physical age of the universe.

This time-scale is remarkably short, 
at least compared to the times we shall shortly discuss.  
It implies that when the universe 
is less than two-hundred times its present age, 
comparable to the lifetime of very low mass stars,
any remaining intelligent life will no longer
be able to obtain new empirical data on the state of large scale 
structure on scales we can now observe.   
Moreover, 
if today $\Lambda$ contributes $70 \%$ of the total energy density 
of a flat ($k=0$) universe, 
then the universe became $\Lambda$-dominated 
at about 1/2 its present age.
The ``in principle" observable region of the Universe 
has been shrinking ever since.  
This loss of content of the observable
universe has not yet become detectable, but it soon will.
Objects more distant than the de Sitter horizon now
will forever remain unobservable.  
On the bright side for astronomers,
funding priorities for cosmological observations 
will become exponentially  more important as time goes on.

\section{The Recoverable Energy Content of the Observable Universe}

As we shall discuss, it will be crucial for the continued
existence of life for the recoverable energy in the universe to be 
maximized.
If the universe is dominated by a cosmological constant,
then although the volume of the universe may be 
infinite
the amount of energy available  to any civilization,
like the amount of information,
is limited to at most what is currently observable, and so is 
finite. 
But what if the cosmological constant is instead zero, or time varying,
so that it does not ultimately dominate the energy
density of the Universe?

Suppose that  at very late times in the history of the universe,
the dominant form of energy density  $\rho_{dom}$ 
scales with the expansion as  $a^{-n_{dom}}$, with $n_{dom}>0$
(if $n_{dom}=0$, then the universe is cosmological constant-dominated).
Equation (\ref{FRWeqns}) can then be solved for the evolution of the
scale factor  ---   $a \propto t^{2/n_{dom}}$.  If $n_{dom}<2$, then
the expansion is accelerating and, as in the case of a 
cosmological
constant dominated universe, 
one is forever limited to the energy and information content 
of a finite subvolume of the universe.
If on the other hand $n_{dom}\geq2$,
then the total energy that can eventually be contained
within the causal horizon may be infinite.

Knowing that 
there are infinite energy reserves ultimately 
containable within the (ever growing) causal
horizon is not enough.  
One must be able to recover the energy to use it!
Can a single civilization 
recover an infinite amount of energy given an infinite 
amount of time in an expanding universe?
The answer, as we now show, appears to be no.

Suppose that intelligent life-forms in the universe
seeking to fuel their civilization
construct machines to prospect and mine the universe for energy.
The energy source they seek to collect may or may not be the dominant
energy density of the universe, 
so its energy density $\rho_{coll}$
can  scale as $a^{-n_{coll}}$,  with $n_{coll}\geq n_{dom}$.
To compete with the decreasing energy density,
the number $N$ of such machines may be increased, 
so at some late time in history let $N \propto t^b$.
The mass $M$ of each machine  may also be changed, 
so that $M\propto t^c$. 
The total collected energy will therefore depend on 
the efficacy ${\cal E}$ of each machine,  
the physical volume per unit time per unit machine mass  
from which the machine is able to extract energy.
Suppose this scales as $t^d$ at late times.
 We allow all the energy recovered
to be funneled into the construction of mining machines,
and ignore the ongoing energy expenditures to 
run the machines. Clearly, this is overly optimistic.  However,
we will find insurmountable difficulties even ignoring this
inevitable energy sink. 

The most optimistic rate of energy recovery is  therefore
\begin{equation}
\Phi = N M {\cal E} \rho \propto t^{b + c + d - 2 n_{coll}/ 
n_{dom}}, 
\end{equation}
while the rate of growth of the total mass of the machines is 
\begin{equation}
\label{NM}
{d\over d t}\left(N M\right) \propto (b+c) t^{b+c-1}. 
\end{equation}
Since the total machine mass can ultimately grow no faster than 
the total recovered energy, we must have  either
\begin{equation}
\label{cequation}
d \geq 2{n_{coll}\over n_{dom}} -1 \geq 1
\quad\quad {\mathrm or}\quad\quad
b + c < 0
\end{equation}
to be able to maintain indefinitely this rate of energy recovery.
If $d \geq 2{n_{coll}\over n_{dom}} -1$, then an infinite amount of
energy can be collected.  However,
if $d <  2n_{coll}/n_{dom} -1$, so that $b+c\leq0$, 
then $\Phi \propto t^p$, with $p<-1$, and 
the total recovered energy will be finite. 
The crucial question is therefore: how fast can the efficacy ${\cal E}$ grow?
The answer depends on the type of energy density that one is collecting.

\subsection{Prospecting for Matter}

First, let us consider prospecting for non-relativistic 
matter ($n_{coll} =3$).  
Because the matter is effectively at rest,
the prospector must bring the matter into the system.
If the prospector makes use only of short range forces 
(those which fall faster than the square of the distance to the machine),
then the prospected volume per unit mass per unit time will saturate,
$d\leq0$. The total recovered energy will be finite.

The prospecting machine would therefore need to use a long range force
to continuously increase its sphere of influence as the universe expands.
The available long range forces (gravity and electromagnetism)
fall off as the inverse square of the distance, 
but grow linearly as the mass (or charge) of the machine.
Using gravity is a  more optimistic option, 
since  the Coulomb force can screened by negative charges.
We therefore consider a massive prospecting machine.
Particles at rest with respect to the comoving expansion, 
if sufficiently close to such an object, will fall towards it.

Simple arguments based on the growth of structure 
imply that the volume of the sphere of influence  of our mining machine
cannot grow as fast as $t$ in an ever  expanding universe.  
Indeed in an ever-expanding universe
all objects have a finite ultimate sphere of gravitational influence.  
Consider a region that has a density $\rho+\delta\rho$ 
which exceeds the mean density $\rho$ of the universe.
If the  region  is sufficiently large, 
gravity will cause the region to expand somewhat more slowly than the average.
The over-density $\delta\rho/\rho$ of the region compared to the mean
will increase.
Once $\delta\rho/\rho$ approaches one, 
the region will decouple from the background expansion, 
grow slightly and then collapse.  

Because there is a uniform background density of material,
the gravitational effect of any local mass distribution becomes 
negligible as one goes to larger volumes -- all objects are
gravitationally influenced only by larger mass overdensities.  
For $n_{dom}\neq 3$ (e.g. curvature, radiation, or cosmological constant
dominated), expansion eventually wins out over collapse on large scales, 
and structure formation ceases;
the gravitationally accessible mass for our ``machine" is therefore
finite.

Only in a matter-dominated ($n_{dom}=3$) flat ($k=0$) universe,
does structure continues to grow hierarchically. 
We do not appear to live in such a universe.
Nevertheless, even in this case 
the gravitationally accessible mass appears to be finite, though  
the ultimate result of large scale structure formation  
would depend upon the spectrum of primordial density perturbations. 

Primordial density perturbations could be absent on large scales, so
that ever larger structures do not form.
In this case the accessible $n_{coll}=3$ energy 
contained within the collapsed perturbations is clearly finite.  
Alternately, non-zero density fluctuations could continue to come inside
the horizon indefinitely.
In this case, structures on ever larger scales will continue to form.
As described above, 
after entering the horizon, 
fluctuations will grow in  size until 
$\delta\rho/\rho\simeq1$.
At this point they will decouple from the expansion and soon begin 
to recollapse.
For the structures we currently observe, 
such as galaxies or cluster of galaxies, 
the recollapse has been (temporarily) halted 
by the internal pressure of the collapsing matter.   
This happened long before their average density exceeded the
critical density of a black hole of that mass, 
$\rho_{BH} = {3 c^2\over 32\pi G_N^3 M^2}$.
However, once the collapsing structures  are sufficiently large 
and  the collapsing matter is sufficiently cold,
there is no known source of internal pressure to halt the collapse
until after they have exceeded this critical density.
Not only is the energy accessible to civilizations finite in such cases, 
but it all must ultimately end in the singularity inside a black hole.
This is identical in detail with the ultimate fate of life in a 
collapsing universe.  Thus, in a flat, matter dominated universe,
life either is stranded on isolated islands of finite total energy, 
or is swept into a large black hole.

Hence, it appears that in any cosmological model,
only a finite amount of $n_{coll} =3$ energy can be recovered by 
static machines.

\subsection{Relativistic matter, and mobile mining machines}

If the energy to be mined involves radiation, rather than matter, then
$n_{coll} =4$.   This applies to a uniform background of radiation,
such as the Cosmic Microwave Background.  If the source of such
radiation lies instead in discrete concentrations of matter, then
the preceding analysis applies, and only a finite total energy can be
mined.

For the case of an $n_{coll} =4$, background,
one must perform a different analysis.   
It is also worth recognizing that we 
can include here the special case in which we move mining
machines to scoop up matter or energy.   The case of a static detector
intercepting radiation  will be equivalent to a moving detector with
$v=c=$constant, for example.

Imagine collectors of  effective area $A$ intercepting the energy
(with $A$ equal to the number of scattering centers times 
the cross section for scattering of each scattering center),  
so that
\be
\Phi = \rho N A v.
\ee
At late times $v\propto t^e$  with $e \leq 0$. 
(For
a static detector receiving radiation with $v=c$, $e=0$.)
Note that a moving  machine will be slowed down as it
sweeps up energy from the background,
requiring a continuing input of energy into the machine.  
As the mass of the machine grows, 
the energy input required will also increase with time. 
We will ignore this need to input kinetic energy for the moment, 
as it is irrelevant for what turns out to be the optimal possibility:
a static detector receiving radiation. 

At first sight,  it seems that
the most efficient collectors would be black holes. 
As a black hole passes through the universe
(or as radiation streams by the black hole),
it effectively traps all material which falls within
the disk spanned by its event horizon.
The  area of the black hole's horizon scales as $M_{BH}^2$, 
so $A \propto t^{2c}$

Equivalently, we might optimistically
consider investing collected photons
in new collecting machines which might somehow coherently convert them
into material particles.  In this case, the cross section for these
machines would grow as the square of the the number of material particles.
(Note that this is the most optimistic assumption one can make.)
In either case, we can then consider a rate of energy collection
optimistically given by:

\begin{equation}
{d E\over d t} = \gamma E^2 t^{-8/n_{dom}}
\end{equation}
with $\gamma = {\cal F}(16\pi G^2/c^3)\Omega^{rad}_o \rho_c
t_o^{8/n_{dom}}$ in flat space ($k=0$). Here ${\cal F}$ is the
gravitational focusing factor, which is a number of order $1$ that
depends  on the velocities of the particles being collected.
(The curved space result is more complicated,
but the final results are unchanged, as we will describe.)
Here $\rho_c$ is the critical density of the universe.
(If $\rho>\rho_c$ then $k>0$; if $\rho<\rho_c$ then $k<0$.)
$\Omega^{rad}_o\rho_c$ is the current energy density in radiation;
$t_o$ is the current age of the universe.
The long term behavior of $E(t)$ in this case is:
\begin{equation}
\lim_{t\to\infty}E(t) = 
\frac{E_o}{1 - {\gamma E_o t_o^{1-8/n}\over 8-n}}  .
\end{equation}
This is finite so long as the initial mass $M_o = E_o/c^2$
is less than a critical  value:
\begin{equation}
M_c \equiv {(8-n)c\over  16\pi G^2 \rho_c t_o \Omega^{rad}_o}
\end{equation}
This critical mass is equal to the mass within the entire visible
universe times a factor of order $1/\Omega^{rad}_o$.  
Since $\Omega^{rad}_o\simeq 10^{-4}$, even under this overly
optimistic assumption,
the radiation energy that such a machine (black hole or otherwise) can
collect is finite.  (For a black hole, we have the additional problem that
the energy collected is stored for a long time, as the black hole
lifetime goes as $M^{-3}$.  Hence the usable power quickly falls in 
this case, so that the power required to run energy metabolizers could
quickly exceed the available supply.).

We can understand this general result as follows.  If such a machine,
say a black hole, could collect infinite energy, this would imply that
the entire visible universe could collapse into such an object.  But
general arguments based on the growth of large scale structure tell
us that only if one starts out with an  extra-horizon-sized black hole can
this be the case.

Next, it is worth pointing out that 
not only the total energy but also the number of photons
received by any individual scattering center, 
integrated over the history of the universe, is finite.
This can be seen by integrating 
the photon number density times the relevant
scattering cross section, over time, as follows
\be
N_{tot} \propto \int_{t_i}^\infty n_{\gamma}  \sigma dt .
\ee
Since $n_{\gamma} \propto t^{-6/n_{dom}}$,  
and since the total
mass of the prospector and thus the number of scattering  centers is finite,
this integral is finite 
unless the electromagnetic cross section rises steeply with decreasing energy.
However, as all such cross sections approach a constant at low energy, 
the number of photons collected is therefore finite.
We shall return to this issue later in this paper.   

Finally, we note that in the case of a cosmological constant-dominated
universe, Gibbons-Hawking radiation exists.  One might imagine that this
radiation, at a constant temperature related to the horizon size, could
provide an energy source to be tapped.  However, while it would take work
to keep any system at a lower temperature (see below), the energy momentum
of this radiation is that appropriate to a cosmological term and not a
standard radiation bath, and thus it cannot be extracted for useful work
without tapping the vacuum energy itself.

\subsection{Extended sources of energy}

For $n_{coll} < 3$, recoverable energy sources are infinitely 
extended objects 
(cosmic strings have $n_{coll}=2$ , domain walls give $n_{coll}=1$)  
which do not fall freely into any localized static machine, 
thus once again $d<1$, and the total collectible energy is finite. 
One caveat to this argument is that we have assumed 
that the energy density to be cannibalized is, on average, 
uniformly distributed throughout space, so that general 
scaling relations for energy density are appropriate.  
An exception to that assumption is any topological defect such
as cosmic strings or domain walls,
in which the number density redshifts as $a^{-3}$,
however the linear/surface energy density of the defect
remains constant so that $\rho$ scales as $a^{-2}$ or $a^{-1}$ 
respectively.  Could the energy in such defects be cannibalized?
The problem is that the rate at which one
can extract energy from the strings (or walls)
is finite (at any given time there is only a finite
amount of string in the observable universe)
and one can not continue extracting the energy indefinitely. 
Why? Because whatever strategy one develops for mining the string, 
the universe can, and will, emulate.
Consider cosmic strings.  If they are unstable,
then their energy density will eventually decline exponentially.
If they are (topologically) stable, then the
only way to mine them is by nucleating either monopole-
antimonopole pairs 
or black-hole pairs along their  length.
However, the universe  will  also avail itself of precisely the 
same strategy.  
In fact, no matter what, black hole pairs will eventually nucleate 
on the strings  
and consume them.  The length of string in the observable universe 
is growing at most as  a power of  time,
whereas at long enough time 
(longer than the characteristic time for a black
hole pair to nucleate on a string)
the rate at which black hole pairs 
are eating the string becomes exponential.
The total length of string which you can eventually
mine may be extremely long, but it must ultimately be finite.
Could the rate of black-hole pair nucleation along the string
itself be a rapidly decreasing function of time?
Only if the gravitational ``constant'' were changing 
appropriately--
a possibility perhaps in some theories of gravity, 
but  hardly a good bet for the ultimate success of life.

\vskip 0.1in

On an optimistic note, while we argue that only finite energy
resources are available, it is worth noting that in all expanding
cosmologies, the actual amount is very large indeed, allowing life-forms
with metabolisms equivalent to our own to exist, in principle for
times in excess of $10^{50}$ years.  Other issues, including proton
decay, for example, may become relevant before an energy crisis
arises.  Nevertheless, we next address the question of whether, even
with finite energy resources, life might, in principle, be eternal.

\section{Living with Finite Energy In an Ever-Cooling Universe}

It was Dyson \cite{dyson} who first seriously addressed the question of 
the ultimate fate of life in an open universe.
Having assumed that the supply of energy ultimately 
available to life would be finite
(as we have shown above always to be the case), 
he realized that life will be
forced eventually to go on an ever stricter diet
to avoid consuming all the available energy.

The first question he identified is
whether consciousness is associated with a specific
matter content,
or rather with some particular structural basis. If the former,
then life would need to be maintained at
its current temperature forever, and could
not be sustained indefinitely with finite resources.
If however consciousness could evolve into whatever
material embodiment best suited its purposes
at that time, ``then a quantitative discussion of the
future of life in the [expanding] universe becomes possible''
\cite{dyson}. We will assume here, for the sake of argument,
that it is structure  which is essential;
we  will also assume that the embodiment of that structure
{\em must} be material.

Dyson assumed a scaling law that is
independent of the particular embodiment that
life might find for itself, as follows:
\vskip 0.1in
{\bf Dyson's ``Biological Scaling Hypothesis (DBSH):}  
If we copy a living creature, quantum state by quantum state, 
so that the Hamiltonian
\be H_c = \lambda U H U^{-1} \ee
(where $H$ is the Hamiltonian of the creature, 
$U$ is a unitary operator, and $\lambda$ is a positive scaling 
factor), 
and if the environment of the creature is similarly copied 
so that the temperatures of the environments of the creature and 
the copy 
are respectively $T$ and $\lambda T$, 
then the copy is alive, subjectively identical to the original 
creature,
with all its vital functions reduced in speed by the same factor 
$\lambda$. " \cite{dyson}
\vskip 0.1in

As Dyson  pointed out, the structure of the Schrodinger equation
makes the form of this scaling hypothesis plausible.
We shall adopt the DBSH here 
and comment later on possible violations.

The first consequence of the DBSH explored by Dyson
is that the appropriate measure of time 
as experienced by a living creature is not physical (i.e. proper) 
time, $t$, 
but  the ``subjective time''
\be
u(t) = f \int_0^t T(t') d t',
\ee
where $T(t)$ is the temperature of the creature
and $f$ is a scale factor with units of $({\rm K sec})^{-1}$
which is introduced to make $u$ dimensionless.
Dyson suggests $f\simeq(300 {\rm K sec})^{-1}$
to reflect that humans operate at approximately $300^oK$
and a ``moment of consciousness'' lasts about one second,
however the precise value is immaterial,
only the fact that $f$ is essentially constant.

The second consequence of the scaling law is that any
creature is characterized by its rate Q 
of entropy production per unit of subjective time.
A human operating at $300K$ dissipates about 200W, therefore 
\be {\rm Q} \simeq 10^{23} \ee 
Dyson asserts that this is a measure 
of the complexity of the molecular structures 
involved in  a single act of  human awareness.
Though one might question whether this
entire Q should be  associated with the act of awareness, 
since in the typical human a significant fraction of  Q is devoted 
to  intellectually non-essential functions,
nevertheless this does suggest that a civilization of conscious 
beings
requires ${\rm log}_2{\rm Q} >  50-100$.

A creature/society with a given Q and temperature $T$
will convert energy to heat at a minimum rate of 
\be
m = k f Q T^2 .
\ee
$m$ is the minimum metabolic rate in ergs per second of physical
(not subjective) time and $k$ is Boltzmann's constant.
It is crucial that the scaling hypothesis
implies that $m \propto T^2$, one factor
of $T$ coming from the relationship between energy and 
entropy,
the other coming from the assumed (isothermal) temperature 
dependence  of the 
rate of  vital processes.

Suppose that life is free to choose its temperature  $T$.
There must still be a physical mechanism for radiating the  
creature's excess heat into the environment.   
Dyson showed that there is an absolute limit
on the rate of disposal of waste heat as 
electromagnetic radiation 
\be
I(T) < 2.84{N_e e^2 \over m_e \hbar^2 c^3} (k T)^3
\ee
where 
$N_e$ is the number of electrons (or positrons) at temperature 
$T$.  
This limit arises from the rate of dipole radiation by the 
electrons. 
Any other form of radiation will have a stronger dependence on 
$T$, 
at least at low $T$:
massless neutrinos are emitted from matter only by weak 
interactions, 
which are mediated by massive  intermediate particles,
gravitational radiation is coupled only to quadrupoles. 
Both therefore scale more strongly with temperature at low 
temperature.
All free particles other than photons, gravitons and neutrinos
are massive thus their emission is exponentially suppressed at low 
temperature.

The rate of energy dissipation, $m$, must not exceed the power 
that can be radiated, if the object is not to heat up, implying a 
fixed lower bound for the
temperatures of living systems:
\be
T > {2 Q \hbar f  \over N_e k 2 \alpha  \gamma} {m_e c^2\over 
k} 
\simeq {Q\over N_e} 10^{-12} {\rm K}.
\label{tminrad}
\ee
$N_e$ cannot be increased without limit,
since the supply of energy (and hence mass) is finite.  
Q however cannot be decreased without limit. 
(A system of  one bit complexity is probably not living, 
a  system of less than one bit complexity is certainly not 
living.)
The slowing down of metabolism
described by the DBSH is therefore insufficient to allow
life to survive indefinitely.

Dyson goes on to suggest a strategy -- hibernation.
Life may metabolize intermittently but continue to 
radiate away waste heat  during hibernation.  
In the active phase, life will be in thermal
contact with the radiator at temperature $T$.
During hibernation, life will be at a lower temperature,
so that metabolism is effectively stopped.
If a society  spends a fraction $g(t)$ of its  physical
time active and  a fraction $[1-g(t)]$ hibernating,
then the total subjective time will be given by
\be
u(t) = f \int_0^t g(t')T(t') d t'
\ee
and the average rate of dissipation of energy
is 
\be
m = k f Q g T^2
\ee
The constraint (\ref{tminrad}) is replaced by
\be
T(t) > T_{min} \equiv {Q\over N_e} g(t) 10^{-12}{\rm K} .
\ee
Life can both keep in step with this limit
and have an infinite subjective lifetime.
For example, if $g(t)= {T(t)\over T_o}$,
with $T_o > (Q/N_e)10^{-12}K$,
and  we let $T(t)$ scale as  $t^{-p}$,
then the total subjective time is 
\be
u(t) \propto \int^t t'^{-2p} d t'
\ee
which diverges for $p\leq1/2$.
The total energy consumed scales as
\be
\int^t m(t')d t' \propto \int^t t'^{-3p}d t'
\ee
which is finite for $p>1/3$.
Thus if $1/3 < p \leq 1/2$,
the total energy consumed is finite and the total subjective time
is infinite.  

It is clear that this strategy will not work in
a cosmological constant-dominated universe.
This is because a cosmological constant dominated universe   is
permeated by background radiation at a constant temperature
$T_{deS} = \sqrt{\Lambda/12\pi^2}$.
A particle detector (such as a radiator for radiating away energy)
will register the de Sitter background radiation,
and bring the radiator into thermal equilibrium with the 
background. (Note however, for the reasons mentioned earlier,
the energy in the cosmological constant cannot be tapped or
converted into useful work if the cosmological
constant remains constant.) 
Therefore $T_{deS}$ is the minimum temperature
at which life can function. 
It is then impossible to have both infinite subjective lifetime
and consume a finite amount of energy.
Life must end, at least in the sense of being forced to have
finite integrated subjective time.
(Note that one cannot use the de Sitter radiation as
a perpetual source of free energy. A cold body will
indeed be warmed by the radiation, but it takes
more free energy to cool the body than can be extracted.)

In fact, we now argue that this hibernation strategy will fail
not only in a cosmological constant dominated universe, 
but in any ever-expanding universe.
In order to implement the hibernation strategy there are two
challenges.  First, 
one must construct alarms which must be relied
on to awaken the sleeping  life.
Second, one must recognize that eventually thermal contact
with one's surroundings effectively ends:  

\vskip 0.1in
\noindent{ 1) A standard alarm clock, one which is subject to
the DBSH, suffers from the same constraints as
those imposed above upon life.  This clock must be powered at
some level to keep
time and it will thus dissipate energy. If it is 
subject to the DBSH, then there is a minimum temperature at which 
it can be operated. 
The alarm clock is a system of some complexity $Q_{alarm}$,
which  as Dyson showed cannot therefore be operated at arbitrarily 
low temperature.  Since $Q_{alarm}$ cannot be reduced
forever, eventually one cannot operate a standard alarm clock.
As we shall show in section 4, even if one could 
manage to expend energy only to wake up the hibernator, 
and not to run the alarm clock in the interim,
the alarm clock would still eventually exhaust the entire store of 
energy. }

\vskip 0.1in
\noindent{ 2)  The  living system is not in thermal equilibrium.
As we have shown, the integrated number of CMB photons 
received over all time is finite.
Therefore, after a certain time 
the probability of  detecting another CBR photon,
integrated over all of future history approaches zero.  
Thus, thermal contact with this background (and all other
backgrounds) is lost.}   

Note also that in any case, the Dyson
expression for dipole radiation, assumed above, clearly breaks down at
some level, notably when the wavelength of thermal radiation becomes very
large compared to the characteristic size of the radiating system.  Put
another way, the thermal energies will eventually become small compared
to the characteristic quantized energy levels of the system, at which
point radiation will by suppressed by a factor $ \approx e^{-
E_{char}/kT}$ compared to the estimate of Dyson.  Once this occurs,
further  cooling will be difficult.  The only alternative to avoid
this is to increase the characteristic size of the system with $a$, which
presents its own challenges.
 
\vskip 0.1in

Lastly, another problem ultimately presents itself independent
of the above roadblocks. Alarm
clocks are eventually guaranteed to 
fail.      In the low temperature mode these failures
may be statistical or quantum mechanical.
If the number of material particles which can be assembled is 
finite,
the catastrophic failure lifetime may be large, 
but it cannot be made arbitrarily so.
In the absence of a sentient being to repair
the broken alarm clock, hibernation would continue forever.

In fact this argument about broken alarm clocks  applies equally
well to living beings themselves.  
Eventually, the probability of a catastrophic
failure induced by quantum mechanical fluctuations
resulting in a loss of  consciousness  becomes important.
One might hope to avoid this fate by keeping the structures in 
contact with their surroundings (which can suppress quantum 
fluctuations
such as tunneling).  However, hibernation requires precisely the
opposite, and moreover, we have seen that such contact gets smaller
over time.  
In any case, for a plethora of reasons, under the DBSH, it appears that
consciousness is eventually lost in any eternally expanding universe.

\section{Beyond the Biological Scaling Hypothesis}

Clearly, if consciousness is to persist indefinitely, one
must consider moving beyond the DBSH.  
The DBSH assumes implicitly
that only rescalings and no fundamental improvements or alterations
can be made in the mechanisms of consciousness.  
A particular consequence is  that the rate of 
entropy production scales as $T^2$.
Can one do better? 
 
It may appear that a full answer to this question requires that
we understand the mechanisms of consciousness.  However, in fact,
our above discussions indirectly point to an approach which
demonstrates that as long as the mechanism of consciousness is
physical, life cannot endure forever. 

Let us return momentarily to the question of whether there are
 non-standard alarm clocks which can be operated
at arbitrarily low temperature, with arbitrarily low energy per 
cycle.  
This possibility hearkens back to recent results on the thermodynamics
of computation, and more importantly, issues of reversible 
quantum computation.  

It was long thought that computation is an entropy generating process,
and thus a heat generating process.   More recently
\cite{feynman,bennett,landauer} it has been pointed out that as long as 
(a) one is
in contact with a heat bath, and (b) one is willing to compute
arbitrarily slowly, then computing itself can be a reversible process.

This opens the possibility that if living systems can 
alter their character so that consciousness can
be reduced to computation, one could in principle reduce the amount of
entropy, and hence the amount of heat produced
per computation arbitrarily, if one is willing to
take arbitrarily long to complete the computation.  Thus, metabolism,
and the continued existence of consciousness,
could violate the DBSH. 

There are two problems. First, as we have
shown, living
things cannot remain in thermal equilibrium with the cosmic 
background forever, so inevitably the process of computation
becomes irreversible.   Also, the question
of computational reversibility is in some sense irrelevant, since
the process of erasing, or resetting registers inevitably produces
entropy.  If one simply reshuffled data back and forth between
registers, reversibility would be adiabatically possible in principle.
However, we have shown that only a finite number of material particles
are accessible.  
Thus any civilization can have only a finite total memory
available, and resetting registers is therefore essential 
for any organism interacting with its environment, 
or initiating new calculations.  
While an existence, even nirvana, might be possible without this, 
we do not believe it is sensible to define this as life.  
Life therefore cannot proceed reversibly, and organisms cannot
continue computationally metabolize energy into heat at less than
essentially
$kT$ per computation.  In this case, one must perform a detailed
analysis to determine if the energy radiated can continue to cool
a system so that its metabolism falls fast enough to allow progressively
less energy utilization, leading to a finite integrated total energy
usage.  We find that the constraints on such radiation even in the most
optimistic case require the density of the radiating system to reduce
along with the expansion.

We do not provide the details of this analysis here  because we believe
there is a more general argument which establishes that consciousness
cannot be eternal. In order to perform any computation, quantum or
classical, at least two states are needed.  One can in principle force the
computation to proceed in one direction or another, reversibly, by
altering adiabatically the external conditions.  However, if erasures are
performed, or if heat is generated because one is not in perfect
equilibrium with the environment, then after the computation one must be
in a lower energy state than before the computation, as heat has been
radiated.  To perform an infinite number of calculations then implies one
must have an infinite tower of states.  This does not require infinite
energy, if the states approach an accumulation point near the ground
state. However, no finite system has such a property.  
(The emission of arbitrarily many massless particles of
ever low energy should not be regarded as adding new states,
since such particles cannot be confined in a finite material system.)
Hence, no finite
system can perform an infinite number of computations.  Thus, if
consciousness can be reduced to computation, life, at least life which
involves more than eternal reshuffling of the same data, cannot
be eternal.  It may be that this reductionist view 
 of consciousness as computation is incorrect.
However, it is hard to imagine a physical basis for consciousness which
avoids the scaling relationships we have described.

Finally, it is worth emphasizing another issue that we have only
peripherally noted thus far.   We have shown that it is impossible to
collect more than a finite amount of any quantity that scales as
$1/a^3$.  However, the entropy density of the universe scales in
this fashion.  Thus, independent of issues of whether there is
infinite information in an infinite universe, it is impossible
to collect more than a finite amount.  Effectively even an
infinite universe allows only
a finite computational system.

\section{Conclusions}

The picture we have painted here is not  optimistic.  
If, as the current evidence suggests, we live in a cosmological 
constant
dominated universe, the boundaries of empirical knowledge will 
continue to decrease
with time.  The universe will become noticeably less observable on 
a time-scale
which is fathomable.   Moreover, in such a universe, 
the days --- either literal or metaphorical --- are numbered
for every civilization. 
More generally,  perhaps surprisingly, we find that eternal 
sentient
material life is implausible in any universe.  The
eternal expansion which Dyson found so  appealing
is a chimera. 

We can take solace from two facts.  The constraints we provide here are
ultimate constraints on eternal life which may be of more philosophical
than practical interset.  The actual time frames of interest which limit
the longevity of civilization on physical grounds, are extremely long,
in excess of $10^{50}-10^{100}$ years, depending upon cosmological and
biological issues. On such time-scales much more pressing issues,
including the death of stars, and the possible ultimate instability of
matter, may determine the evolution of life.  

Next, and perhaps more important, 
strong gravitational effects on the geometry or topology of the universe
might effectively allow life, or information, 
to propagate across apparent causal boundaries, 
or otherwise obviate the global spatial constraints we claim here.  
For example, it might one day possible to  manipulates such effects
to artificially create baby-universes 
via wormholes or black hole formation or 
via the collision of monopoles \cite{tanmaymark}. 
Then one might hope that in such
baby universes conscious life could eventually appear, 
or that one might be able to move an arbitrarily 
large amount of information into or out of
small or distant regions of the universe. 
While these are interesting possibilities, 
at this point they are vastly more speculative than 
the other possibilities we have discussed here.

\section*{Acknowledgements}

We thank C. Fuchs, H. Mathur, J. Peebles, C. Taylor, M. Trodden 
and T. Vachaspati  for discussions and J. Peebles and 
F. Wilczek for comments on the
manuscript. This research is supported in part by the DOE and NSF.


\begin{thebibliography}{99}
\baselineskip=15pt

\bibitem{perl}
Perlmutter, S., {\it et.al.}, LBNL-42230, Dec 1998., astro-ph/9812473
and references therein.

\bibitem{kirsh}Riess, A.G., {\it et. al.} Submitted to Astron.J., astro-ph/9805201
and references therein.

\bibitem{kraussturner}Krauss, L.M., and Turner, M.S., 
{\it J. Gen. Rel. Grav.} {\bf 27}, 1137 (1995).

\bibitem{ostriker}Ostriker, J.P., and Steinhardt, P., 
{\it Nature} {\bf 377}, 600 (1995).

\bibitem{krauss}Krauss, L.M., {\it Ap. J.} {\bf 501}, 461 (1998).

\bibitem{dyson} Dyson, F.J., {\it Rev. Mod. Phys.}  {\bf 51}, 447 (1979).

\bibitem{feynman} Feynman, R.F., {\it Feynman lectures on computation}, 
ed.  Hey, A.J.G.  and Allen, R.W., Addison-Wesley, Reading, Mass, 1996.

\bibitem{bennett}Bennett, C.H., {\it Int. J. Theor. Phys.}, {\bf 21}, 905
(1982)

\bibitem{landauer} Landauer, R., {\it Physics Today}, May 1991, 23

\bibitem{tanmaymark} Borde, A., Trodden, M., and  Vachaspati, T.,  
CWRU-P25-98, Aug 1998., gr-qc/9808069.

\end{thebibliography}
\end{document}